\documentstyle[preprint,aps,psfig]{revtex}
\begin{document}
\draft
%\psdraft
\preprint{LBNL-40497}
\title{A microscopic calculation of secondary Drell-Yan
production in heavy ion collisions
}
\author{C.~Spieles, L.~Gerland, N.~Hammon, M.~Bleicher, 
S.A.~Bass, H.~St\"ocker, W.~Greiner}
\address{Institut f\"ur
Theoretische Physik,  J.~W.~Goethe-Universit\"at,
60054 Frankfurt
am Main, Germany}
\author{C.~Louren\c{c}o}
\address{CERN PPE, CH-1211 Geneva 23, Switzerland}
\author{R.~Vogt}
\address{Nuclear Science Division, LBNL, Berkeley, CA 94720, USA\\
Physics Department, University of California, Davis, CA 95616, USA}
\maketitle
\vspace*{-0.9cm}
\begin{abstract}
  A study of secondary Drell-Yan production in nuclear collisions is
  presented for SPS energies.  In 
  addition to the lepton pairs produced in 
  the initial collisions of the projectile and target nucleons, we
  consider the potentially high dilepton yield from hard valence
  antiquarks in produced mesons and antibaryons.
  We calculate the secondary Drell-Yan
  contributions taking the collision spectrum of hadrons from the
  microscopic model URQMD.  The contributions from meson-baryon
  interactions, small in hadron-nucleus interactions, are
  found to be substantial in nucleus-nucleus collisions at low
  dilepton masses. Preresonance collisions of partons may further 
  increase the yields.
\end{abstract}

\section{Introduction}

A unique experimental signature from the early stage of violent heavy
ion collisions is provided by dilepton radiation. The (thermal)
radiation of leptons is highest when the energy density has reached a
maximum. Since leptons are insensitive to strong interactions, the
signal is not distorted during the further evolution of the
system.  Recent reports on intermediate~\cite{EXP1} and high
mass~\cite{EXP2} muon pairs produced in heavy ion collisions have
attracted much attention.  The suppression of the J/$\psi$ signal, as
extracted from NA38/NA50 measurements, might indicate that
deconfinement has been reached in central Pb+Pb reactions~\cite{KhSa}.
In the mass region below the $\rho$-meson, on the other hand, the
measured $\rm e^+e^-$ pair spectrum is enhanced relative to the
expected `cocktail' of hadronic decays~\cite{ceres}.  A good
understanding of the continuum dilepton spectrum, in the relevant
invariant mass window, is a prerequisite for the correct
interpretation of the data.

Rate estimates of Drell-Yan pair production~\cite{Dre70} have been
performed systematically for different systems, energies, and
invariant masses~\cite{Vo92}. They are essentially based on
extrapolations of p+A reactions where a linear scaling with A is
observed for the Drell-Yan production cross section~\cite{Al90}: \[
\sigma_{\rm pA}= \sigma_{\rm 0}~\rm A \, . \] This linear scaling can be
understood in the Glauber picture of the hadron-nucleus cross section,
constructed at high energies using the AGK cutting
rules~\cite{AGK73}. (For a simple dynamical picture of this effect,
see~\cite{Far75}.)

A number of different theoretical
frameworks~\cite{Kap92,Kae92,Dum93,Wi95,Ge93,Gav96} have been used to
explore the degree to which the hot matter produced in ultrarelativistic
heavy ion collisions contributes to the total dilepton radiation.
Dilepton emission from thermal sources such as a quark-gluon plasma or hadron
gas, whether in or out of equilibrium, cannot be neglected
relative to nucleon-nucleon contributions to the Drell-Yan process for masses
below the J/$\psi$ peak --- at least at higher energies than those presently
available.  Another source of dileptons which has not been fully explored
is Drell-Yan production by interactions between produced, or secondary,
hadrons.  Unless these secondary dilepton sources are understood, the thermal
contributions cannot be reliably extracted.   In this paper, we study
possible refinements to Drell-Yan production in nucleus-nucleus collisions
from secondary hadron interactions at current fixed-target energies.
It is shown that these secondary collisions can serve as an important source
of $m\sim 2$~GeV dileptons due to the availability of valence antiquarks
in mesons and antibaryons. 
\section{The model}

In order to investigate secondary dilepton production at SPS
energies\,\footnote{At RHIC and LHC energies the heavy flavour (c and
  b) combinatorial decays are expected to dominate the dilepton
  continuum \cite{Gav96}.  Moreover, the nuclear parton distribution functions
  at such small $x$ values are still quite uncertain.  Therefore, we will
  not report any high energy calculations in this paper.}  a
microscopic hadronic transport code, URQMD~\cite{UQMD}, is employed to
obtain a realistic collision spectrum of secondary hadrons.  The
differential Drell-Yan cross section is computed at leading order (LO)
using the standard equation~\cite{Ha78}:
\begin{eqnarray}
  \frac{{\rm d}^2\sigma}{{\rm d}m^2{\rm d}y}({\rm AB}\rightarrow l\bar l X)=
  \frac{4\pi\alpha^2}{9m^2s}\sum_q e_q^2 \left[
  q^{\rm A}(x_{\rm A},m^2) {\overline q}^{\rm B}(x_{\rm B},m^2) +
  {\overline q}^{\rm A}(x_{\rm A},m^2) q^{\rm B}(x_{\rm B},m^2)\right] \; ,
\end{eqnarray}
where $q(x,m^2)$ and $\overline q(x,m^2)$ denote the quark and antiquark
densities, $\sqrt{s}$ is the center-of-mass energy
of the colliding hadrons, $m$ is the invariant mass of the
lepton pair, $x_{\rm A}=\sqrt{\tau}~e^y$ and $x_{\rm B}=\sqrt{\tau}~e^{-y}$
with $\tau=m^2/s$, and $y$ is the dilepton rapidity in the \emph{cms}
frame.  The convolution of the parton distributions is weighted by the
square of the charge of the annihilating quarks, $e_q$.  Since the
calculation is performed at leading order only, the result is
multiplied by a ``K-factor''~\cite{Gav95}.  Although the ratio between the
calculated next-to-leading order (NLO) and LO Drell-Yan cross sections is a
function of $m^2$ and $y$, the dependence is small, $\sim 25$\,\% for $m>1.5$
GeV and $0<y<1$ at $\sqrt{s} = 19.4$ GeV. Thus, for simplicity, we have used a
constant K factor.  This is reasonable considering the accuracy of the present
experimental data which does not reveal any strong mass
dependence of the experimental K factor\,\footnote{This situation might
  change in the near future when the very high statistics
  Drell-Yan differential cross sections of E866~\cite{866} become available.}
 (the ratio between the measured data and the LO calculation).

According to the URQMD hadronic model, most of the interacting
particles in the early stage of
the reaction are baryon and meson resonances.
In this study, if two hadrons have the same quark content, their parton
distributions are assumed to be equivalent, e.g. the proton parton
distributions are used for the $N^*(1440)$ and the pion parton distributions
are used for the $\rho$.  We have employed the GRV
LO parametrizations of the parton distribution functions of
nucleons~\cite{GRV95} and pions~\cite{GRV92a}, since they provide a
consistent set for both species (same $\Lambda_{\rm QCD}$ and similar
initial scale $\mu_0$).  The lepton pair production cross section is
calculated for each hadron-hadron collision and weighted by the
inverse of the total hadron-hadron cross section. The distributions
from these elementary hh-collisions are then summed.

\begin{figure}[p]
\vspace*{\fill}
\centerline{\psfig{figure=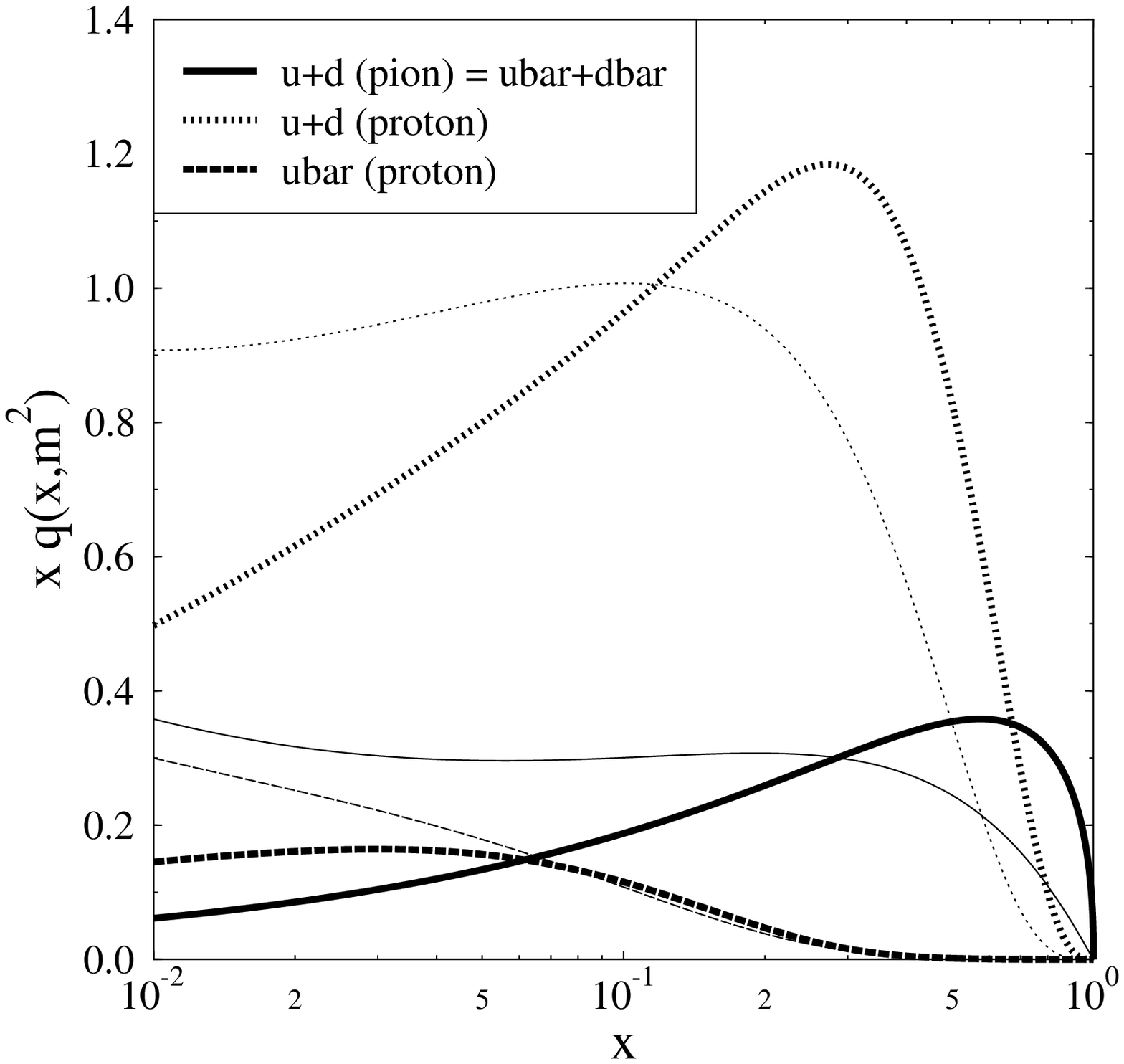,width=16cm}}
\caption[]{The GRV LO parton distribution functions of protons~\cite{GRV95}
  and pions~\cite{GRV92a} for valence and sea quarks. We show the values
  for ($u+d$) quarks in the proton (dotted) and in the pion (full
  lines), and the values for $\overline u$ quarks in the proton (dashed
  lines). The thick lines indicate the distributions for $m^2\approx
  0.3\, \rm GeV^2$, the thin lines those for $m^2=9\, \rm GeV^2$.
  Note the hard valence antiquark distribution in the pion. The valence
  antiquark distributions in the antibaryon correspond to the valence
  quark distributions in the baryons. 
\label{struc}}
\vspace*{\fill}
\end{figure}

It is clear that in pion-nucleon, pion-pion and antinucleon-nucleon
collisions, \emph{valence} quark-antiquark annihilation can play a significant
role in the Drell-Yan process. Figure~\ref{struc} shows the
parton distribution functions of protons~\cite{GRV95} and pions~\cite{GRV92a}
from the GRV~LO parametrization, for two different values of
$m^2$. For $x\stackrel{>}{\sim}0.05$ the probability to find an
antiquark is much higher in a pion than in a nucleon.  The pion-nucleon cross
sections are consequently higher than the
nucleon-nucleon cross sections, especially when $m/\sqrt{s} \stackrel{>}{\sim}
0.1$. (See~\cite{Ha78}.)
On the other hand, in heavy ion collisions the typical meson-baryon
and antibaryon-baryon interactions are certainly less energetic than
the primary nucleon-nucleon collisions.

While the usual Drell-Yan calculations of nucleus-nucleus collisions only
consider interactions of the initial nucleons, in the present
scenario four different processes may contribute:
baryon-baryon, meson-baryon (and meson-antibaryon), meson-meson\,
\footnote{The Drell-Yan process in
meson-meson interactions should not be confused with
processes such as $\pi^+\pi^-\to l^+l^-$ where the whole meson is involved in
the annihilation rather than a single parton. In any case, dileptons from pion 
annihilations are not expected to make an important contribution to the 
intermediate mass region we are concerned with here.} and
baryon-antibaryon interactions.
Thus, it is interesting to estimate the relative importance of these
contributions to the total Drell-Yan cross section.

The typical scaling of the Drell-Yan cross section with target mass
observed in hadron-nucleus collisions, accounted for here, 
assumes an equal probability for
Drell-Yan pair production in each nucleon-nucleon interaction.
In addition, in  a separate microscopic simulation the meson-baryon,
meson-meson and baryon-antibaryon collisions, involving newly
produced hadrons, are calculated within the URQMD transport model. In order
to conserve energy and momentum, the
energy loss of baryons which have already interacted is taken into
account in these {\em secondary} interactions.

For simplicity, the produced hadrons follow straight-line trajectories
across the reaction zone without energy loss ($0^\circ$-scattering,
only elastic interactions).  The center of mass energy of these
collisions is taken directly from the fully dynamical evolution of all
hadrons within the URQMD simulation. In the usual cascade
prescription, leading hadrons, containing valence quarks from the
incident nucleon, can interact within their formation time with a
reduced but finite cross section. Since the constituent
\emph{antiquark} content is much more relevant for the Drell-Yan
process, we only consider collisions of formed hadrons.  The formation
time of produced hadrons, on average $\tau_F\approx 1$~fm/c, is
determined by the string fragmentation prescription used in
URQMD~\cite{UQMD} (defined as the first crossing point of the quark
and antiquark or quark and diquark which form the hadron).  The exact
formation time depends on the mass and momentum of the produced
hadron.
By imposing a minimum hadron formation time, we effectively 
neglect any possible annihilation of produced partons which have not
yet hadronized (``preresonance'' scattering).
%The introduction of a formation time precludes interactions of
%produced partons which have not yet hadronized (``preresonance''
%scattering). 
In Sec.~\ref{primordials} we return to this problem.

\section{The effect of a flavour asymmetric sea}
%%%%%%%% Teil von Nils

Important consequences for the sea quark distributions have recently been
discussed: modifications at small $x$ and the violation of the Gottfried
sum rule~\cite{gott}. Two effects are primarily important at low $x$.  The HERA
data has shown a steep rise in the proton structure function $F_2$ 
for $x\stackrel{<}{\sim}
0.01$ \cite{HERA}.  Additionally, the sea quark and gluon distributions can be
significantly modified in nuclear targets \cite{Arn}.  The observed low $x$
shadowing and the low $x$ rise in the parton sea distributions are not relevant
for SPS energies where $0.077 < x < 0.52$ for $1.5 < m < 10$ GeV at
midrapidity.  At higher energies, such as those at RHIC and LHC, these effects
cannot be neglected.  However, as we now describe, the violation of the
Gottfried sum rule has important consequences at the SPS energies.

The Gottfried sum
rule defines the difference between the proton and neutron structure
functions~\cite{gott}
\begin{eqnarray}
I_{GSR}&=& \Sigma (0,1) = \int_{0}^{1} \frac{dx}{x} (F^{p}_{2} -F^{n}_{2})
\nonumber\\
&=&\frac{1}{3} \int_{0}^{1} dx (u_v - d_v)+\frac{2}{3} \int_{0}^{1}dx
(\overline{u}- \overline{d}) = \frac {1}{3}~~ {\rm if}~~
\overline{u}=\overline{d}.
\end{eqnarray}
Since the neutron distributions are not directly measured, $F_2^n$ is obtained
assuming that $u^p(x,m^2)=d^n(x,m^2)$ and
${\bar u}^p(x,m^2)={\bar d}^n(x,m^2)$.
All global structure function analyses prior to 1992 assumed
that the light-quark sea distributions are flavour independent,
$\overline{u}(x,m^2)=\overline{d}(x,m^2)$.
On the contrary, based on their
$F^{n}_{2}/F^{p}_{2}$ measurements, NMC found that~\cite{NMC}
\begin{eqnarray}
\Sigma (0.004,0.8)=\int_{0.004}^{0.8} \frac{dx}{x} (F^{p}_{2} - F^{n}_{2})
=0.227\pm0.007~~ ({\rm stat.}) \pm 0.014~~({\rm syst.})
\end{eqnarray}
at $m^2 =4~{\rm GeV}^2$.  A straightforward comparison implies that
$\overline{d}>\overline{u}$. These results have been confirmed by p+p and p+d
asymmetry studies of Drell-Yan production by NA51 \cite{NA51} 
and by preliminary
results from E866 \cite{866}.  The lack of a Regge $f-A_2$ exchange
degeneracy also leads one to expect that \cite{MRSD}
\begin{equation}
\Delta \equiv \overline{d}-\overline{u} \sim x^{-\alpha _R}
\end{equation}
at small $x$, where the Regge intercept is $\alpha _R \approx 0.5$.

\begin{figure}[p]
\vspace*{\fill}
\centerline{\psfig{figure=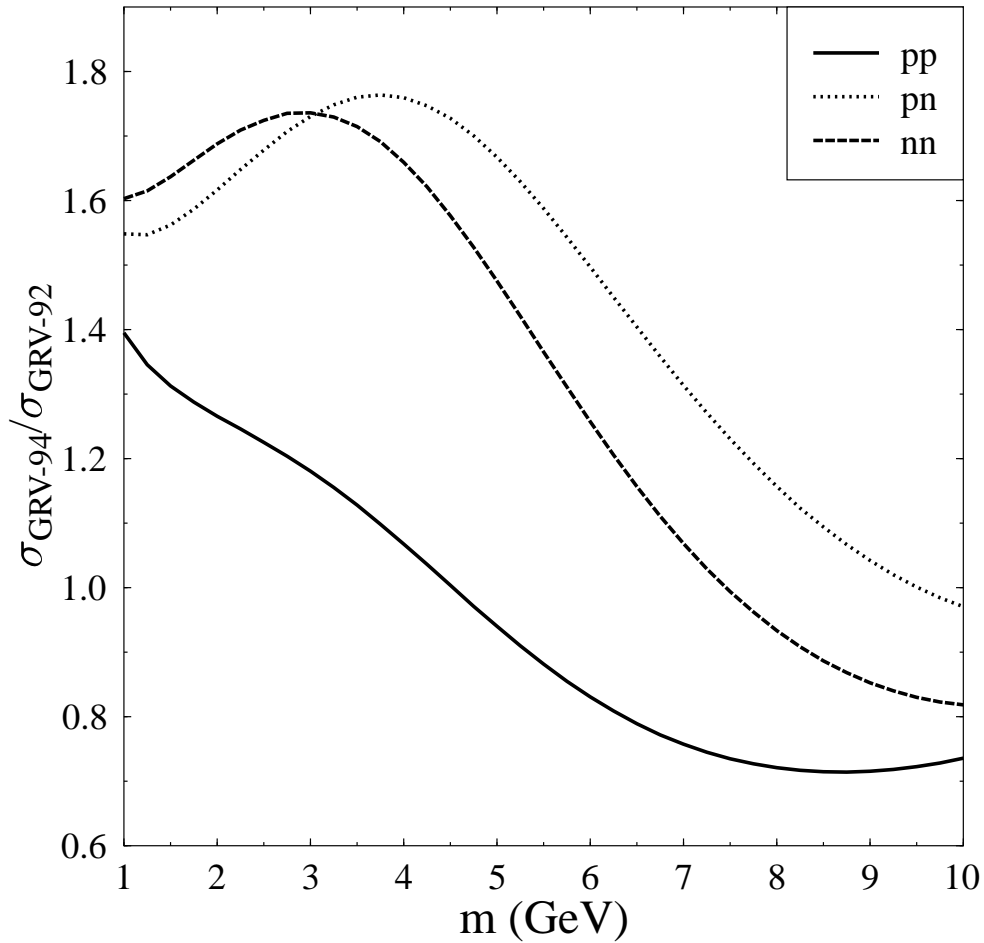,width=16cm}}
\caption[]{Drell-Yan cross section calculated with the recent GRV~94~LO
parametrizations~\cite{GRV95} where $\bar u \neq \bar d$, relative
to the calculation done with the GRV~92~LO set~\cite{GRV92b} which
assumes a symmetric sea. Notice that the K factor to be applied to 
these calculations is 2 for GRV LO 92 and 1.5 to GRV LO 94.
%The ratio of the Drell-Yan cross section from the GRV~94~LO
%  parametrizations of~\cite{GRV95}, where $\bar u \neq \bar d$ to 
%  the GRV~92 result for a symmetric sea~\cite{GRV92b}.
\label{9592}}
\vspace*{\fill}
\end{figure}

With this in mind, we compare two versions of the GRV proton parametrizations,
the flavour-symmetric set from 1992~\cite{GRV92b} and the 1994
version~\cite{GRV95} that
explicitly takes into account the NA51 results by choosing
\begin{equation}
\overline{u}=\frac{1}{2} \left[ (\overline{u}+\overline{d})- \Delta \right],~~~~
\overline{d}=\frac{1}{2} \left[ (\overline{u}+\overline{d})+ \Delta \right]
\end{equation}
Another factor contributing to the difference between the two sets
 is the starting distribution used to obtain the two GRV valence
distributions: the flavour symmetric KMRS B- \cite{KMRS} for GRV 92 and the 
flavour asymmetric MRS A \cite{MRSA} for GRV 94.  These starting distributions 
differ at low $x$  and in the fitted value of $\Lambda_{\rm QCD}$ as well as 
in their treatment of the $\overline u$ and $\overline d$ distributions.
To study the difference between these two GRV parametrizations,
we evaluate the lepton pair production cross sections for p+p,
p+n and n+n interactions.  (Note that at $y=0$, p+n = n+p.)
The ratios of cross sections from the GRV 94 relative to the GRV 92
parametrizations are shown in Fig.~\ref{9592} as functions of $m$
at $y=0$ and $\sqrt{s} =15$ GeV.  The effect of the flavour asymmetric sea,
most pronounced for p+n and n+n interactions, is clearly important.
Inclusion of the flavour asymmetric sea decreases the isospin dependence of the
Drell-Yan cross section.  The isospin correction needed for Pb+Pb interactions
is 1.3 with the GRV 92 LO set \cite{EXP2} and 1.03 with the GRV 94 LO set.  The smaller
isospin correction is also found when MRS G distributions \cite{MRSG} are used.

Since neither corresponding $\overline d - \overline u$ data nor
parton distributions with a flavour asymmetric sea exist for pions, we use the
flavour symmetric GRV distributions for the pion \cite{GRV92a} in our meson
interactions.  One might
suppose the asymmetry to be less pronounced for the pion since the $\pi^+ (u
\overline d)$ and $\pi^- (\overline u d)$ each contain a $u$ and a $d$ quark or
antiquark.

\section{Proton-nucleus and pion-nucleus reactions}

\begin{figure}[p]
\vspace*{\fill}
\centerline{\psfig{figure=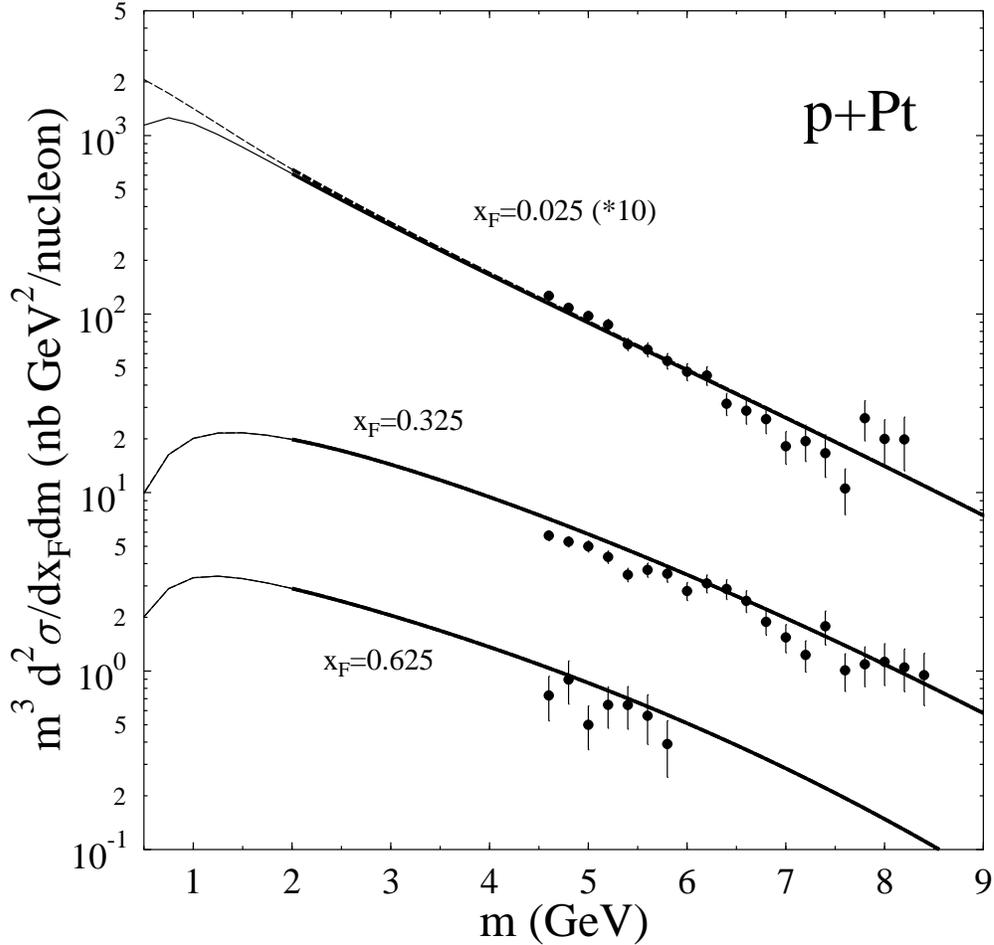,width=16cm}}
\caption[]{Dilepton mass distributions from p+Pt collisions at $p_{\rm
    lab}=400$~GeV/c calculated for three $x_F$ values.  The
  contribution of baryon-baryon collisions is shown (solid line)
  separately from the total spectra (dashed line) which includes
  secondaries.  The data points are from experiment
  NA3\protect\cite{na3}.  We use a K factor of 1.5 for NN collisions
  (calculated with the GRV 94 nucleon PDFs) and of 2 for pion induced
  reactions (calculated with the GRV 92 pion PDFs).  We use thinner
  lines for $m<2$~GeV to emphasize the relatively large theoretical
   uncertainty in the calculation.
%  Since perturbative Drell-Yan production is generally assumed to hold
%  for $m>2$ GeV, we use thin lines for $m<2$ GeV to indicate the
%  theoretical uncertainty in the calculation.
\label{na3}}
\vspace*{\fill}
\end{figure}

Figure~\ref{na3} shows the mass spectrum of dimuons in p+Pt collisions
at $p_{\rm lab}=400$~GeV/c for three different longitudinal momentum
windows.  Our calculation with the GRV 94 parton distributions
provides a reasonable
description of the NA3 data~\cite{na3} with $\rm K=1.5$.  In addition
we have compared the model calculation with p+Cu collisions at
$p_{\rm{lab}}=800$~GeV/c~\cite{e605} and found equally good
agreement.  In fact, the Drell-Yan mass and $x_F$ double differential cross
sections measured by NA3 \cite{na3}, E605 \cite{e605} and E772 \cite{e772}
experiments, are all in good agreement with each other, as shown in
Ref.~\cite{patmcgqm96}.

The proton-nucleus calculations reveal that the contribution from
secondary collisions is less than 10\,\% of the dilepton yield for
masses $m>1.5$~GeV.  The relative importance of secondary
contributions decreases with increasing $x_F$. At $x_F=0.325$ and
$x_F=0.625$, the original Glauber type calculation is increased by
less than 1\,\%, even at the lowest masses.

\begin{figure}[p]
\vspace*{\fill}
\centerline{\psfig{figure=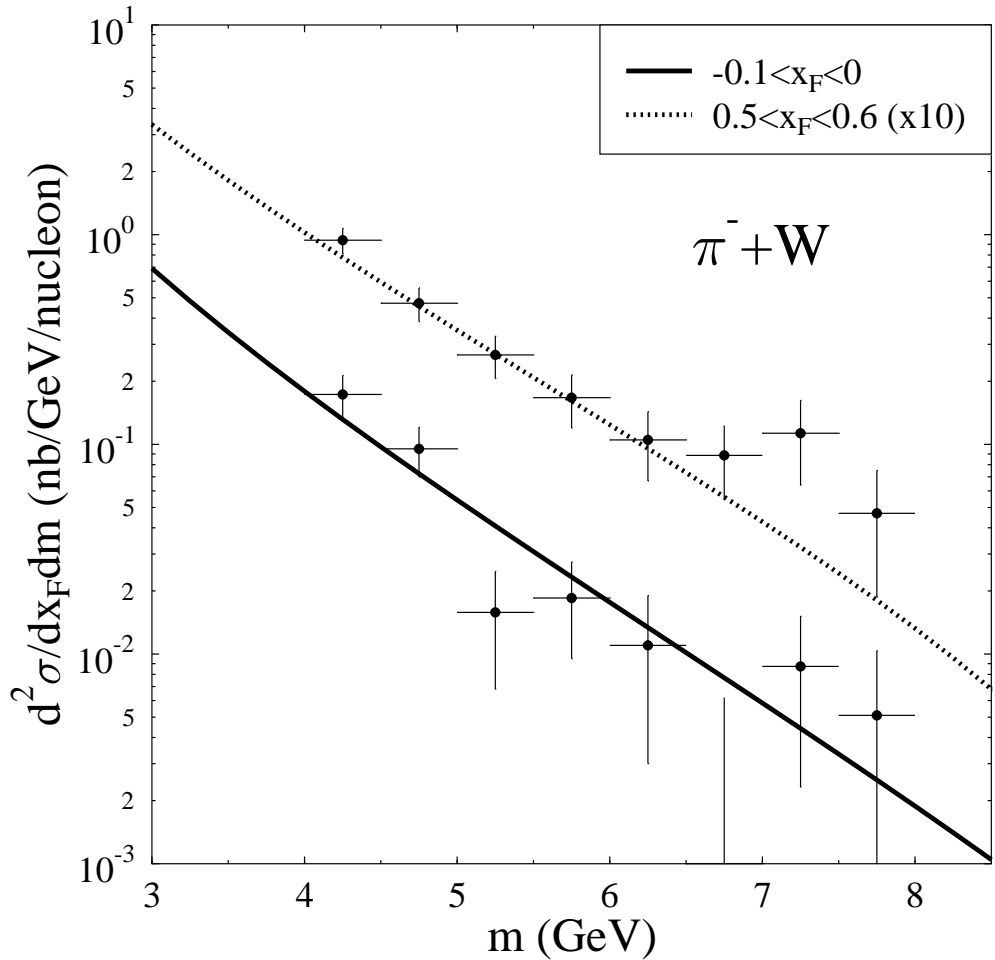,width=16cm}}
\caption[]{Mass spectrum of dileptons for two $x_F$ bins
  in $\pi^-+W$ collisions at $p_{\rm lab}=125$~GeV/c.  We show our
  leading order calculations with data from experiment
  FNAL-537\protect\cite{expFNAL537}.  We use a K factor of 2 for these
  pion induced reactions.
\label{fnal537}}
\vspace*{\fill}
\end{figure}

The dimuon spectrum produced in $\pi^-+W$ reactions, at
$p_{\rm{lab}}=125$~GeV/c, as measured by the E537
experiment~\cite{expFNAL537}, is presented in Fig.~\ref{fnal537} for
a central and a forward $x_F$ bin.  The
results of our leading order calculations, scaled by a K
factor of 2, are shown. A different K factor relative to the value used in
the proton-induced reactions is needed since we use the GRV 92 pion set
with the GRV 94 proton set.  
Although the measurements have large
uncertainties, the agreement between our calculations and the data
supports the validity of our treatment of secondary meson-baryon
collisions.

\section{nucleus-nucleus collisions}

\begin{figure}[p]
\vspace*{\fill}
\centerline{\hspace*{-0.5cm}\psfig{figure=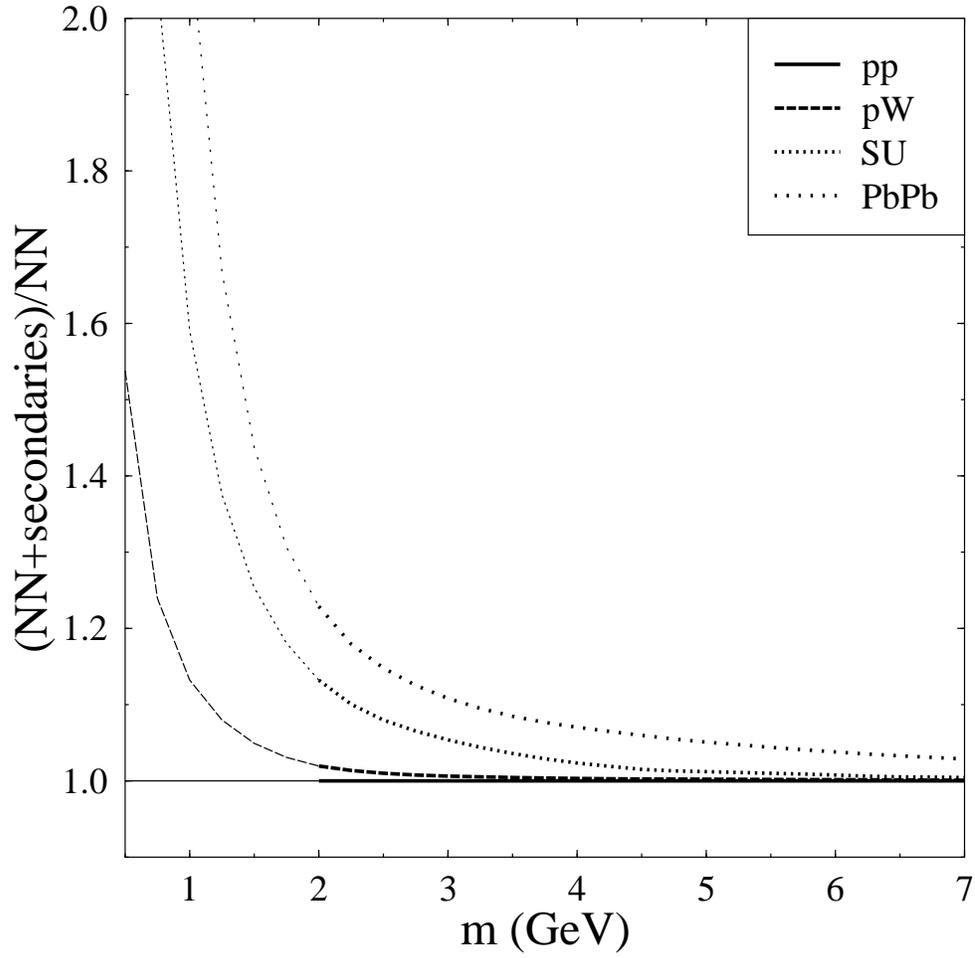,width=16cm}}
\caption[]{Ratio between the total dilepton mass spectrum and the
  nucleon-nucleon scattering alone, at $y_{\rm cms}=0.5$,
  for different systems: p+W and S+U at 200~$A$~GeV, Pb+Pb at 160~$A$~GeV.
\label{sec}}
\vspace*{\fill}
\end{figure}

Figure~\ref{sec} shows the relative importance of the secondary
dilepton production, compared to nucleon-nucleon collisions alone,
for different projectile/target combinations.  In the mass region
around 2~GeV the additional contributions are not negligible in
nucleus-nucleus collisions while the increase is less than 5\,\% in
the p+W case.
For masses above 3~GeV, the additional production beyond nucleon-nucleon
collisions from our hadronic calculation is less than 10\,\% in both S+U and
Pb+Pb interactions.  Evidently the high mass dilepton yield is not
significantly enhanced by the relatively low energy
secondary scatterings.  At masses below 2~GeV, on the contrary, it
appears that the contribution from collisions of newly produced
particles cannot be neglected.  At masses around 1.5~GeV, an enhancement
of 25\,\% is expected in S+U interactions and 45\,\% in Pb+Pb interactions.
Clearly, the importance of secondary contributions to
the Drell-Yan cross section increases with the system size.  
Note, however, that this enhancement is insufficient to account
for the discrepancy reported in Ref.~\cite{EXP1} between the
data and the ``expected sources''.

Although the perturbative QCD calculation of dilepton production at
2~GeV and below can only be used as a qualitative guideline, it is clear
that the low mass dilepton continuum in nuclear collisions can be
enhanced relative to a linear superposition of
nucleon-nucleon interactions.  At least part of the observed
enhancement of muon pairs at intermediate and low masses~\cite{EXP1}
might be caused by this previously neglected hadronic source.  
The perturbative calculation is definitely unreliable for
masses below 1~GeV where a substantial increase in electron pair production
in nucleus-nucleus collisions has been observed~\cite{ceres}.
To emphasize the theoretical uncertainty of the perturbative calculation
for masses below 2~GeV, in Fig.~\ref{sec} and in the following figures,
we use thinner lines in that region.
%To draw attention to the uncertainty of the perturbative calculation
%for masses below 2~GeV, in Fig.~\ref{sec} and the following figures,
%we use thinner curves in this region.

\begin{figure}[p]
\vspace*{\fill}
\centerline{\hspace*{-0.5cm}\psfig{figure=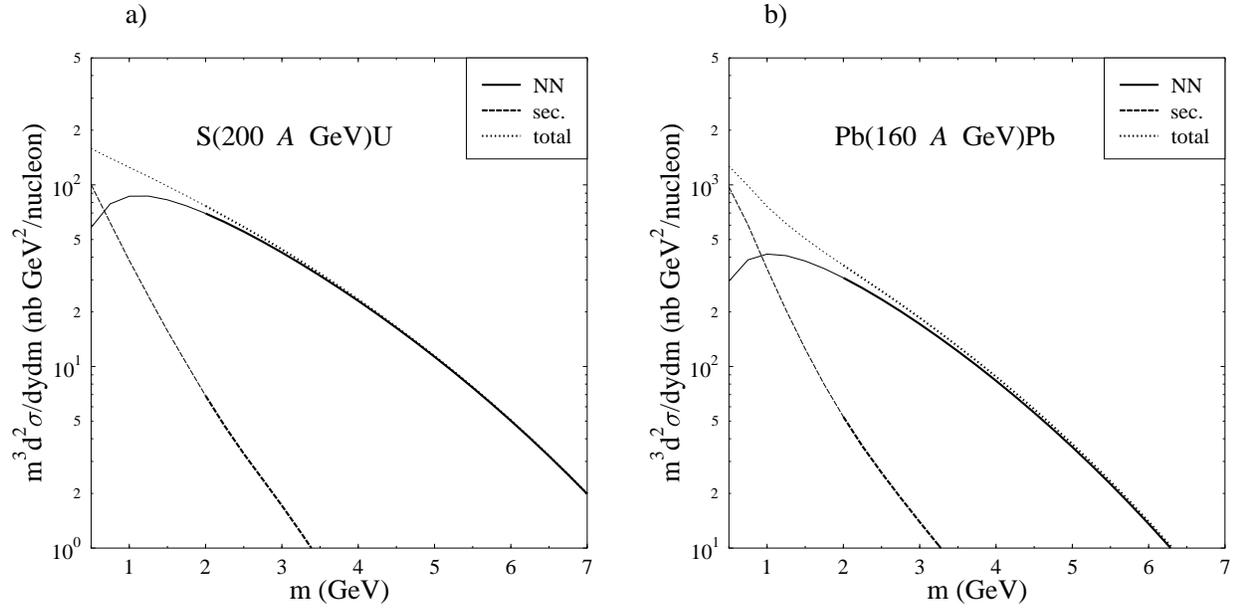,width=17cm}}
\caption[]{Differential Drell-Yan cross section, per unit rapidity, at
$y_{\rm cms}=0.5$, for the 10\,\% most central collisions 
of S+U at $p_{\rm lab}=200$~$A$~GeV (left) and of Pb+Pb 
at $p_{\rm lab}=160$~$A$~GeV (right). The different
Drell-Yan contributions from the hadronic simulation are shown.
\label{fig6}}
\vspace*{\fill}
\end{figure}

The relative contributions to the
calculated dilepton mass spectra at $y_{\rm{cms}}=0.5$ in central
S+U and Pb+Pb collisions are presented in Fig.~\ref{fig6}.
The decomposition of the secondary dilepton yield into each channel
(meson-baryon, meson-meson, baryon-antibaryon) shows that by far the most
important secondary contribution is from
meson-baryon interactions.  For masses above $\sim 3$~GeV they are
virtually the only source of secondary dileptons. At midrapidity and for
masses $\sim 1$~GeV, they account
for at least 90\,\% of the secondary yield while meson-meson collisions
contribute $\sim 10$\,\% and baryon-antibaryon collisions
contribute less than 1\,\%.

\section{Primordial sources of lepton pairs and formation time}
\label{primordials}
The standard Drell-Yan process corresponds to the interaction of fully
formed hadrons.  However, it was shown \cite{KhSa}
that, during the early stages of the system evolution, partons
can scatter and annihilate before they have come on mass-shell\,
\footnote{There are some similarities between the off-shell interactions of the
primordial partons and the soft parton annihilation model proposed to explain
the ``anomalous'' low mass lepton pairs \cite{pisut}.}.  
Thus, we postulate the
existence of another dilepton source: the annihilation
of quarks and antiquarks which are not yet bound in a
color-singlet hadron.

To estimate the importance of these ``primordial'' or ``preresonance'' 
$q\bar q$ annihilations, we have calculated the contribution of such
processes assuming that the asymptotic parton distribution functions
are also valid for the primordial states.  This assumption is distinct
from thermal dilepton production where the quark and antiquark are in
a thermal environment, in or out of equilibrium, with temperature-dependent
parton densities.
We thus relax the restriction that the partons can only interact after they
have hadronized.  This is done very simply in our
microscopic cascade calculation by decreasing the formation time of
the produced hadrons, $\tau_F$, from the ``default'' value of around 1~fm/$c$.

\begin{figure}[p]
\vspace*{\fill}
\centerline{\hspace*{-0.5cm}\psfig{figure=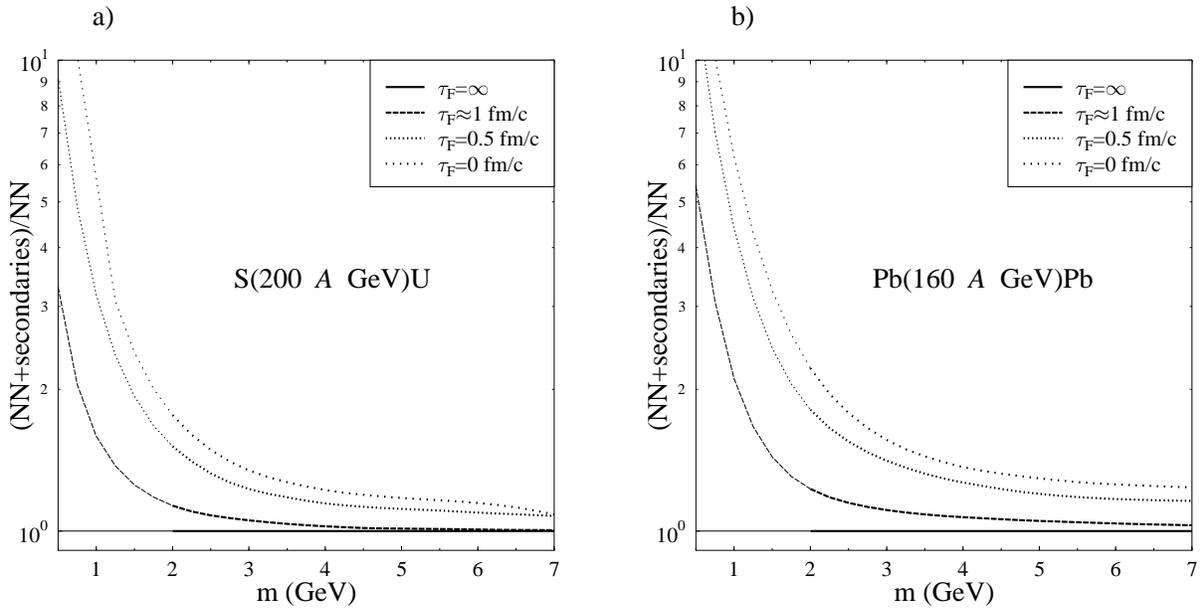,width=17cm}}
\caption[]{Ratio between the total dilepton mass spectrum and the
nucleon-nucleon scattering alone, at $y_{\rm cms}=0.5$, for 
S+U colisions at $p_{\rm lab}=200$~$A$~GeV (left) and for Pb+Pb 
collisions at $p_{\rm lab}=160$~$A$~GeV (right).
The calculations correspond to different
formation times for the secondary hadrons.
\label{prim}}
\vspace*{\fill}
\end{figure}

The importance of this primordial (preresonance) contribution to the dilepton 
mass spectra is shown in Fig.~\ref{prim}.  We see that when
we set $\tau_F=0$, the secondary sources of dileptons become much more
important than previously indicated.  Indeed, the
secondary dilepton yield increases by a factor of $\sim
5$ at all masses compared to the calculations with the default
$\tau_F\approx 1$~fm/$c$.  These primordial interactions also affect p+A
reactions, enhancing the secondary contribution.  For example,
%although 
the dilepton yield at $m=4.5$~GeV and $x_F=0.025$ shown in
Fig.~\ref{na3} increases by 25\,\% when $\tau_F = 0$.  Nevertheless,
good agreement with the NA3 data is retained.

In Fig.~\ref{prim} we have also included a calculation with an
intermediate formation time for the produced hadrons, $\tau_F=0.5$~fm/c,
in order to study the sensitivity of the calculations to this
(model dependent) parameter.  As we might expect, the reduced
average formation time leads to an increased dilepton yield due to
the larger number of possible hadron-hadron interactions. With this value of
the formation time, the enhancement in the range $1.5<m<2$ GeV shows quite good
agreement with the intermediate mass data \cite{EXP1}.
On the other hand, an
infinite formation time, $\tau_F=\infty$, corresponds to the exclusion
of all secondary interactions, in agreement with the Glauber-type
Drell-Yan expectation.  A recent work~\cite{tform} has already
pointed out the strong influence of the hadron formation time on
dilepton production via $\pi^+\pi^-$ annihilation.

\section{Conclusion}

Our study indicates that the lepton pair continuum
cross section for masses up to 3~GeV should not be interpreted
solely in terms of Drell-Yan production in nucleon-nucleon interactions,
even at SPS
energies. The yield from secondary contributions depends strongly on
the rapidity region and, of course, on the dilepton mass. Preresonance
interactions are estimated to enhance the secondary yield by up 
to a factor of 5.

Because meson-baryon interactions dominate secondary lepton pair production,
the relative importance of secondary dilepton sources evidently 
increases linearly with the number of produced pions at these energies.
It may thus be possible to distinguish the secondary yield from thermal 
dilepton sources which are expected to scale quadratically with the pion
multiplicity. The multiplicity dependence also indicates that 
the dilepton yield from scattering of secondaries is likely to be considerably
higher at RHIC and LHC energies.

%\narrowtext

\acknowledgements 
We would like to acknowledge interesting discussions with J.~Schukraft.
We thank the Institute for Nuclear Theory at the University of
Washington, Seattle, where this work was initiated, for its kind
hospitality.
This work was supported by Graduiertenkolleg, Gesellschaft f\"ur
Schwerionenforschung, Bundesministerium f\"ur Bildung und Forschung, Deutsche
Forschungsgemeinschaft. The work of R. V. was supported
in part by the Director, Office of Energy Research, Division of Nuclear Physics
of the Office of High Energy and Nuclear Physics of the U. S.
Department of Energy under Contract No. DE-AC03-76SF0098.

\clearpage

\clearpage

\clearpage

\clearpage

\clearpage

\clearpage

\clearpage

\clearpage

\end{document}